\documentclass[notoc,12pt]{JHEP3}
\usepackage{amsmath,amssymb}
\input{epsf}
\newcommand{\startappendix}{
\setcounter{section}{0}
\renewcommand{\thesection}{\Alph{section}}}
\newcommand{\Appendix}[1]{
\refstepcounter{section}
\begin{flushleft}
{\large\bf Appendix \thesection: #1}
\end{flushleft}}
\usepackage{epsfig}

\def\N{{\cal N}}

\def\Dbarslash{\,\,{\raise.15ex\hbox{/}\mkern-12mu {\bar\D}}}
\def\Dslash{\,\,{\raise.15ex\hbox{/}\mkern-12mu \D}}
\def\delslash{\,\,{\raise.15ex\hbox{/}\mkern-9mu \partial}}
\def\delbarslash{\,\,{\raise.15ex\hbox{/}\mkern-9mu {\bar\partial}}}

\newcommand{\EQ}[1]{\begin{equation} #1 \end{equation}}
\newcommand{\AL}[1]{\begin{subequations}\begin{align} #1
\end{align}\end{subequations}}
\newcommand{\SP}[1]{\begin{equation}\begin{split} #1
\end{split}\end{equation}}

\title{Spacetime Virasoro algebra from strings on zero radius
  $AdS_3$}
\author{Paul de Medeiros $^{a}$ and S.~Prem~Kumar $^{a,b}$\\

$^{a}$ Department of Physics, University of Wales Swansea,\\ Singleton
Park,
Swansea, SA2 8PP, U.K.\\\\
$^{b}$ Department of Applied Mathematics and Theoretical Physics,\\
Wilberforce Road, Cambridge CB3 0WA, U.K.\\

E-mail: {\tt p.de.medeiros@swan.ac.uk , s.p.kumar@swan.ac.uk}}
\preprint{SWAT-382, DAMTP-2003-99}
\abstract{ We study bosonic string theory in the light-cone gauge on
$AdS_3$
  spacetime with zero radius of curvature (in string units)
  $R/ {\sqrt{\alpha^\prime}} =0$. We find that the worldsheet theory
  admits an infinite number of conserved
  quantities which are naturally interpreted as spacetime
  charges and which form a representation of (two commuting copies of)
  a Virasoro algebra. Near the boundary of $AdS_3$ these charges are
  found to be isomorphic to the infinite set of asymptotic Killing
  vectors of $AdS_3$ found originally by Brown and Henneaux. In
  addition to the spacetime Virasoro algebra, there is a worldsheet
  Virasoro algebra that generates
   diffeomorphisms of the spatial coordinate of the string
  worldsheet. We find that if the worldsheet Virasoro
  algebra has a central extension then the spacetime
  Virasoro algebra acquires a central extension via a mechanism
  similar to that encountered in the context of the $SL(2,{\mathbb R})$
WZW
  model.
  Our observations are consistent with a recently
  proposed duality between bosonic strings on zero radius $AdS_{d+1}$
  and free field theory in $d$ dimensions.}

\begin{document}
\section{Introduction}

The purpose of this paper is to study certain aspects of string theory on
$AdS_3$ spacetime with zero radius of curvature (in string units).
In certain contexts, string theories on infinitely curved $AdS$ spaces
have been conjectured to be dual to free field theories
\cite{karch1,karch2,dhar,surya}.
Such a duality is expected to occur as a limiting case (of zero 't Hooft
coupling) in the classic example of the holographic correspondence
between type IIB superstring theory on
$AdS_5\times S^5$ and the $SU(N)$ $\N=4$ supersymmetric Yang-Mills theory
in four-dimensional Minkowski space \cite{maldacena,review}\footnote{There
the 't Hooft coupling $\lambda$ of the gauge theory is related to the
radius of curvature $R$ of the $AdS$ background via
$R^2=\alpha^\prime\sqrt\lambda$ whilst $1/N$ controls the string
genus expansion. Pushing Maldacena's conjecture to the limit, we would
then conclude that at $\lambda=0$ (and $N\rightarrow\infty$) the IIB
theory on $AdS_5\times S^5$ with vanishing radius should be dual to
the free $\N=4$ theory.}. Some evidence in support of this
expectation has been presented in \cite{dhar,bianchi}.

It has been further argued in \cite{karch1,karch2} that
(noninteracting) bosonic closed
string theory on $AdS_{d+1}$ with zero radius of curvature is dual to
the free scalar field theory of $N\times N$ matrices (with
$N\rightarrow\infty$) in $d$ spacetime dimensions. In particular, it
was shown that the
Hamiltonian and states of the string theory in light-cone gauge
at $R^2/\alpha^\prime=0$ naturally map to the corresponding
objects in the sector of single-trace states of the free matrix-valued
scalar field theory. Unlike the case of the $\N=4$ theory it is
not clear if this correspondence can be extended to an
interacting field theory and this appears to be related to the fact that
we
cannot treat $R^2/\alpha^\prime$ as a small non-zero
parameter in the string theory. Nevertheless, the proposal of
\cite{karch1,karch2} raises certain intriguing questions -- even at the
level of the free theory. One such question that forms the motivation
for this note occurs in the $d=2$ case wherein we would expect
bosonic string theory on $AdS_3$ in the singular limit to be dual to
the free boson theory in two dimensions. Free field
theory in two dimensions is special in that it has an
infinite-dimensional conformal symmetry generated by two copies of the
Virasoro algebra with a central
extension. The question is whether bosonic
closed string theory on
zero radius $AdS_3$ can also realise (two
commuting copies of) a
{\em spacetime} Virasoro algebra with a non-zero central extension,
which could then potentially be identified as the conformal algebra of a
holographic dual field theory. We find that this is indeed the case.

This question can also be posed in the context of type IIB strings on
$AdS_3\times S^3\times T^4$ with $R$-$R$ three-form background, which
is the near horizon limit of the D1-D5 brane system. In the limit of zero
$AdS$ radius ($R/\sqrt{\alpha^\prime}\rightarrow 0$) one expects
that this
theory is likely dual to the orbifold point of the CFT of the symmetric
product space $(T^4)^k/S_k$ (where $k=Q_1Q_5$ for $Q_1$ D1-branes and
$Q_5$ D5-branes). The light-cone
superstring Hamiltonian in this background has been derived in
\cite{metsaev1, metsaev2}.  The analysis in our paper applies to the
bosonic sector of this theory in the tensionless limit
\footnote{The interpretation of the tensionless limit in the S-dual
  setting, namely the F1-NS5 brane system is not clear to us. In the
  context of the Wess-Zumino-Witten (WZW) model description of the
  theory, this limit  
  has been discussed recently in \cite{Lindstrom:2003mg}.}.

What makes our analysis possible is the well-known
fact that, in light-cone gauge, the classical string action on $AdS$
space (in Poincar\'{e} coordinates) with $R^2/\alpha^\prime=0$ undergoes a
drastic
simplification
\cite{tseytlin, karch1,karch2,dhar}. In particular,
spatial gradient terms on the worldsheet drop out and the string
breaks up into noninteracting bits. In this sense the string indeed
appears to be tensionless, which agrees with a naive interpretation of
the $R^2/\alpha^\prime\rightarrow0$ limit as taking $\alpha^\prime$ (the
inverse tension) to infinity whilst keeping $R$ fixed.
We are able to
show that this theory satisfies
the basic requirement for consistency, namely that the spacetime
isometry algebra of $AdS_3$, realised on the light-cone worldsheet,
closes in the quantum theory.

One of our main observations is that the tensionless string on $AdS_3$
admits an infinite set of exactly conserved quantities (constructed
from worldsheet fields) which satisfy a
Virasoro algebra and which can naturally be interpreted as spacetime
charges.
The interpretation of these conserved quantities as spacetime
charges follows from two facts. First, the $SO(2,2)\cong
SL(2,{\mathbb R})\times SL(2,{\mathbb R})$
isometry group of $AdS_3$ is a global symmetry of the
worldsheet theory and the corresponding Noether charges are in fact
a subset of the infinite set of conserved quantities we
consider. Second, in an asymptotic sense, these conserved
quantities are isomorphic to the
Brown-Henneaux Killing vectors
which generate the asymptotic isometry group of $AdS_3$
\cite{review,brown}. It should be pointed out that the Brown-Henneaux
isometries are not symmetries of the string action. In fact we find
that our conserved quantities can be interpreted as
the Noether charges associated with certain global symmetries of the
string action in light-cone gauge. These symmetries are generated by
global (nonlinear) field redefinitions which match up with the
Brown-Henneaux transformations only in an asymptotic sense.

Furthermore we find in the first quantised string theory a
mechanism for inducing a spacetime central charge which is
similar to the mechanism encountered in the $SL(2,{\mathbb R})$ WZW
model in \cite{Giveon:1998ns}.
In particular the worldsheet theory of the closed tensionless string
in light-cone gauge has a residual invariance generated by
diffeomorphisms of the circle (the spatial coordinate of the closed
string worldsheet), whose generators form a
{\em worldsheet} Virasoro algebra. We find that if this worldsheet algebra
is allowed to have a central extension, it naturally induces a central
term in the spacetime algebra due to the \lq winding' of certain
time-independent closed string fields.
A very similar mechanism was found in
\cite{Giveon:1998ns} in the context of the $SL(2,{\mathbb R})$ WZW model
wherein the spacetime algebra can be constructed from the affine
algebra on the worldsheet and a non-zero spacetime central extension
appears
due to winding of long string states in $AdS_3$.
The existence of a spacetime Virasoro algebra is encouraging as it is
in accord with the expectation that a
theory of gravity on $AdS_3$ should (at least asymptotically) have an
enhanced symmetry
generated by
two commuting Virasoro algebras. The observations above are consistent
with
the expectation that the tensionless string on $AdS_3$ should be dual to a
conformal
field theory in two dimensions.

The outline of our paper is as follows. In Section 2 we briefly review
the classical light-cone gauge action for strings on $AdS$ space and
the limit of zero $AdS$ radius. We
also discuss canonical quantisation of the zero radius system and the
presence of
a worldsheet Virasoro algebra. In Section 3 we construct a set of
conserved
quantities on the worldsheet of the tensionless string and we show
that they are to be identified with the generators of a spacetime
Virasoro algebra. In Section 4, we present our conclusions and
discussions.
Finally, in the Appendix we study issues pertaining to operator
ordering and the closure of the algebra in the quantum theory. We
also point out the existence of (worldsheet realisations of) spacetime
Virasoro algebras for $AdS$ spaces of arbitrary dimension.

\section{Bosonic light-cone strings on AdS}

\subsection{The classical theory}
In this section we briefly review some aspects of bosonic strings on
$AdS$ space in the light-cone gauge. Detailed discussions of this
topic can be found in
\cite{karch2,polchinski,metsaev1} \footnote{A covariant formulation of
tensionless strings on $AdS$ space has been discussed recently in 
\cite{Bonelli:2003zu}. The tensionless string in curved space and
perturbations about that limit have also been studied in detail in
\cite{Lousto:1996hg}}. 
 We begin by considering classical
string
propagation on the patch of $AdS_{3}$ covered by the Poincar\'{e}
coordinates $(x,t,z)$. In these coordinates, the $AdS_{3}$ metric is
\EQ{ds^2={R^2\over z^2}(2dx^+dx^-+dz^2),\label{poincare}}
where $x^+$ and $x^-$ are the light-cone variables defined by
$x^\pm:=(x\pm
t)/\sqrt 2$. The radial coordinate $z$ is non-negative and $z=0$
corresponds to
the boundary of $AdS_3$. On the string worldsheet, parametrised by the
coordinates $(\tau,\sigma)$, we can make the standard
light-cone gauge choice
\EQ{x^+\equiv {x+t\over\sqrt 2}=p_-\tau,}
where the constant $p_-$ will subsequently be identified with the
light-cone momentum. Unlike in flat space, we do not have the freedom to
choose a conformal
gauge for the worldsheet metric $h_{ij}$. Instead we can only choose
worldsheet reparametrisations to set $h_{01}=0$.
With these gauge
choices, the string action is
\EQ{S_{LC} =\int d^2\sigma {R^2\over 4\pi\alpha^\prime
      z^2}\left[-\sqrt{-h_{11}\over h_{00}}(2 p_- \,\dot{x}^- +\dot{z}^2)+
\sqrt{-h_{00}\over h_{11}}(z^\prime)^2\right],}
where $\dot z:= \partial_\tau z$ and $z^\prime := \partial_\sigma z$.
The canonical momentum conjugate to $x^-$ is $\pi_-(\sigma)=
-p_-\sqrt{-h_{11}\over h_{00}}R^2/2\pi\alpha^\prime z^2$ and is a
function of $\sigma$ only. Residual reparametrisations of the form
$\sigma\rightarrow \tilde\sigma(\sigma)$ can then be used to set
$\pi_-(\sigma)=p_-$. This corresponds to a uniform distribution of the
light-cone momentum on the worldsheet. The light-cone string action on
$AdS_3$ is then\footnote{We remark that our convention differs
  slightly from that used in \cite{karch2,metsaev1}. The latter would
  have led to an action $\int p_-{\dot {z}}^2 /2$ for the transverse
  mode. However, this does not alter any of the conclusions in this
note.}
\EQ{S_{LC}=\int d\tau\int_0^1 d\sigma\left[ p_- \,\dot{x}^- +{1\over
2}{\dot
      z}^2- {1\over 8\pi^2\alpha^{\prime 2}}{R^4 \over z^4
    } z^{\prime\, 2} \right].\label{lc1}}
As pointed out in \cite{karch2}, when compared with the light-cone action
for
strings in a flat background, Eq.(\ref{lc1}) (or the corresponding
Hamiltonian) can naturally be
interpreted as the action for a string with variable
tension $R^2/z^2(2\pi\alpha^\prime)$. With this form of the
action we can now freely set the dimensionless parameter
$R^2/\alpha^\prime=0$ and treat it as our definition of the
classical light-cone string
action on zero radius $AdS_3$ or equivalently bosonic tensionless
strings on $AdS_3$:
\EQ{S_{LC}=\int d\tau\int_0^1 d\sigma \left[ p_- \, \dot x^- +{1\over
2}{\dot
      z}^2\right]; \quad\quad{R^2/\alpha^\prime=0}.\label{lc2}}
Note that all of the above can be simply extended to general
$AdS_d$ backgrounds by replacing $z$ with $z^i$ ($i=1,\ldots ,d-2$) the
$d-2$ transverse
coordinates. For the sake of clarity however, we have restricted
ourselves to $AdS_3$
as this is the subject of the note. The light-cone action must, of
course, be supplemented with two Virasoro constraints that follow by
varying the action with respect to the two independent components of
the worldsheet
metric. As usual, these constrain $x^-(\sigma,\tau)$ such that, except for
its zero mode, it is completely determined in
terms of the transverse degrees of freedom (in this case
$z(\sigma,\tau)$). In the singular limit, the constraints are
\EQ{p_- \, \dot{x}^- + {1\over 2}\;\dot{z}^2\; =\; 0,\label{vc1}}
\EQ{p_- \, x^{-\prime} + \dot z\,z^\prime \; =\; 0.\label{vc2}}
In this limit, the above light-cone action Eq.\eqref{lc2} is
remarkably
simple. Since the $\sigma$-derivatives drop out the string breaks up
into free \lq bits' and the action for the field
$z(\sigma,\tau)$ simply describes an infinite number of free
non-relativistic particles. However, the stringiness
of the system is inherent in the constraints above.
It is also clear from
Eq.(\ref{lc1}) why a small
non-zero value for $R^2/\alpha^\prime$ cannot obviously be treated as a
perturbation -- the terms involving this small parameter become singular
near the boundary
$z\rightarrow 0$. Therefore the interpretation of the theory is clear
only at zero $AdS$ radius. Although the
latter leads to a trivial action Eq.\eqref{lc2}, when supplemented with
the
constraints Eqs.\eqref{vc1} and \eqref{vc2}, we will see that it
exhibits certain features expected in a theory of gravity on
$AdS_3$. It is also interesting to note that the light-cone action for
the tensionless string on $AdS$ spaces coincides precisely with the
so-called null string on flat space (see e.g.
\cite{Gamboa:1989px,Isberg:1993av}).

\subsection{Canonical quantization}
From the action Eq.\eqref{lc2} and the constraints
Eqs.\eqref{vc1} and \eqref{vc2}, it follows that both the fields
$z(\sigma,\tau)$ and
$x^-(\sigma,\tau)$ satisfy the free particle equations
\EQ{\ddot{z}(\sigma,\tau)=0,\qquad \ddot{x}^-(\sigma,\tau)=0 \label{eom}}
and therefore grow linearly with time. In order to perform canonical
quantization it is convenient to expand the $\sigma$-dependence of the
worldsheet fields in terms of their Fourier modes on the closed string:
\EQ{z(\sigma,\tau)=\sum_{n=-\infty}^{\infty}(z_n+p_n\tau)e^{2\pi i
    n\sigma};\qquad z_{-n}=z_n^\dagger,\;p_{-n}=p_n^\dagger,}
which is real and periodic. Periodicity of $z$, however, does not
guarantee the
periodicity of the constrained variable
$x^-$ on the closed string. Indeed the general solution of the
constraints
Eqs.\eqref{vc1} and \eqref{vc2} and the equations of motion\label{eom}
leads to the mode decomposition
\EQ{\qquad
x^-(\sigma,\tau)=\sum_{n=-\infty}^{\infty}(x_n^-+p_n^-\tau)e^{2\pi i
    n\sigma}+P_\sigma\,\sigma,}
where
\EQ{x_n^- (n\neq 0)=\sum_{m=-\infty}^\infty{1\over p_-}\;\Big({m\over
    n}-1\Big)z_{n-m}p_m,\qquad P_\sigma= 2\pi i\sum_{n=-\infty}^\infty
{1\over
    p_-}n\;z_ {-n} p_n.\label{xmode}}
The formula for $p_n^-$ can also be deduced from the constraint
Eq.\eqref{vc1} (see e.g. \cite{karch2}).
Clearly the coefficient of the linear term $P_\sigma$ does not
automatically come
out to be zero. It is the momentum flow in the $\sigma$-direction on
the worldsheet.
In fact, in the quantum theory, the vanishing of $P_\sigma$ must be
treated
as a
constraint to be imposed on
physical states of the closed string, such that
\EQ{P_\sigma|\rm {phys}\rangle =0,\label{physical}}
which is equivalent to the condition that $(x^-(1)-x^-(0))$ annihilates
physical
states.
In the context of the ordinary
tensionful string in flat space the analogous requirement leads to the
level matching condition.

With the exception of its zero mode $x_0^-$, the field $x^-$ is
completely determined in terms of the radial variable $z$ of $AdS_{3}$,
via the constraint equations \eqref{vc1} and \eqref{vc2}. Therefore
$x_0^-$ and $z(\sigma,\tau)$ are the two dynamical variables satisfying
the canonical (equal time) commutation relations
\EQ{[x_0^- ,p_-]=i, \qquad [z(\sigma,\tau),\dot{z}(\sigma^\prime,\tau)] =
  i\delta(\sigma-\sigma^\prime). }
The second commutator implies that the Fourier modes of $z$ satisfy
\EQ{[z_m, p_n]=i\delta_{m+n,0},}
with all other commutators vanishing\footnote{Our expressions for
  $x_n^-$ and the canonical commutators differ
  from those in \cite{karch2,dhar} by a factor of $1/p_-$. This can be
traced back to the normalisation of the action. The two
conventions are related by a $p_-$-dependent rescaling of the modes
$z_n$ and $p_n$.
All final results are independent of choice of convention.}. The
canonical momenta conjugate to the coordinates $x^+$ and $z$ are
\EQ{\pi_+=\dot x^-=-{1\over 2p_-}{\dot z}^2,\qquad \pi_z=\dot
  z.\label{momenta}}
The light-cone action and
Hamiltonian for the transverse (radial) coordinate are respectively,
\EQ{S_{LC}=\int d^2 \sigma \, {1\over 2}{\dot z}^2,\quad\quad H_{LC}=
{1\over 2}\sum_{n=-\infty}^\infty p_n p_{-n}\label{ham}}
which, of course, describe non-relativistic free particles.

The canonical commutation relations above must be appended with a
consistent definition of quantum operators. This is clear in the case
of the operator $P_\sigma$ in Eq.\eqref{xmode} which has an ordering
ambiguity. In ordinary tensionful string theory where the worldsheet
is a theory of oscillators there is a natural ordering prescription
for quantum operators, namely normal ordering. Here however, the
worldsheet theory
is not a theory of oscillators and we need to consider
the possible consistent definitions of the quantum operators. The
light-cone Hamiltonian Eq.\eqref{ham} has no ordering ambiguities. We
discuss some of these ordering issues in Appendix A.
\subsection{Worldsheet Virasoro algebra}

Interestingly, as in the case of ordinary tensile string theory, in the
tensionless theory the Fourier modes $y_n$ of the field
$p_-\;x^{-\prime}$ actually constitute a Virasoro algebra,
\SP{\\
[y_m, y_n ]= (m-n)\;y_{m+n}+(c m^3+c^\prime
m)\;\delta_{m+n,0}\label{vir1}}
where
\EQ{y_n=-i\sum_{m=-\infty}^\infty(n-m)z_{n-m}p_m .}
Note that we have allowed
for the most general central extension to the worldsheet algebra. The
values of $c$ and $c^\prime$ depend delicately on the ordering
of quantum
operators. It has been observed in earlier works, in the context of
tensionless strings in flat space (null strings) \cite{Gamboa:1989px},
that depending on the ordering prescriptions (and the related definition
of the
vacuum) $c$ and $c^\prime$ can be non-zero. We review some of
these details in
Appendix A. In what follows we assume
that $c$ has some non-zero value and $c^\prime=0$ and observe the
consequences of this
assumption. This choice is motivated by the ordering prescriptions
introduced in \cite{Gamboa:1989px} which we discuss in Appendix
A\footnote{Following a normal ordering prescription it also turns out
  that $c=2$. Note that $c^\prime$ can be set to zero by a suitable
  choice of an ordering constant in $y_0$ (which is equal to
  $p_-P_\sigma/2\pi$).}.

At this point we find it convenient to introduce a new worldsheet variable
$X(\sigma)$, defined as
\EQ{X(\sigma)\;:= \;x^-(\sigma,\tau)+{1\over
    2p_-}z\dot{z}(\sigma,\tau).\label{bigx}}
The main property of this field $X(\sigma)$
is that it is conserved by virtue of the constraints Eqs.\eqref{vc1}
and \eqref{vc2}. $X(\sigma)$ is time-independent at every point on the
string and encodes an infinite number of conserved quantities on the
worldsheet. In addition, the spatial derivative of this field satisfies
\EQ{p_- X^\prime ={1\over 2}(\dot{z}^\prime z-\dot{z}z^\prime)}
and so its Fourier modes $l_n$ are found to also satisfy a
Virasoro algebra
\EQ{[l_m,l_n]=(m-n)\;l_{m+n}-{c\over 2}m^3\delta_{m+n,0},\label{vir2}}
where
\EQ{l_n=-{i\over 2}\sum_{m=-\infty}^\infty(n-2m)z_{n-m}p_m\;}
and $c$ has the same value as in Eq.\eqref{vir1}. One can understand
the appearance of this Virasoro algebra as a consequence of a residual
invariance of the gauge-fixed action generated by reparametrisations
of the circle. In fact the light-cone action Eq.\eqref{ham} is
invariant under the infinitesimal transformations $\delta z = z^\prime
f(\sigma)+zf^\prime(\sigma)/2$ which can be thought of as a combination of
a
reparametrisation $\delta\sigma= f(\sigma)$ and a \lq spacetime' conformal
transformation. $\{ l_n \}$ are the conserved Noether
charges associated with this invariance and satisfy the algebra generated
by
infinitesimal diffeomorphisms of the circle.

With all the above properties to hand, a straightforward calculation shows
that $X(\sigma)$ obeys the following
commutation relation
\SP{[X(\sigma),X(\sigma^\prime)]=&\;{i\over
    2p_-}\, \left( X^\prime(\sigma)+X^\prime(\sigma^\prime) \right)
\left( \theta(\sigma-\sigma^\prime)-\theta(\sigma^\prime-\sigma)
\right)\\\\
&+{i\over p_-}\, \left( \sigma^\prime X^\prime(\sigma^\prime)-\sigma
X^\prime(\sigma) \right) -{1\over 2\pi i}\,{c\over
  2p_-^2}\, \delta^\prime(\sigma-\sigma^\prime).\label{xcom}}
The time-independent field $X(\sigma)$ and its commutators form
the basis for our subsequent spacetime
interpretation of the worldsheet theory at zero $AdS$ radius.
\section{A spacetime Virasoro algebra}

On general grounds \cite{brown}, a theory of gravity on
(asymptotically) $AdS_3$ spacetimes is expected to have an
infinite-dimensional
symmetry group generated by two commuting copies of a Virasoro
algebra, thus leading to an interpretation as a spacetime conformal
field theory in two dimesions. This holographic interpretation arises from
the fact
that although gravity in three dimensions does not have any
propagating degrees of freedom, it has so-called global degrees of freedom
corresponding to the degrees of freedom of the graviton that can be
gauge transformed to the $AdS_{3}$ boundary. In the context of string
theory on
$AdS_3\times S^3 \times M^4$ with non-zero $NS$-$NS$ 2-form flux (the
$SL(2,{\mathbb R})$ WZW model), this phenomenon has been understood in a
series of beautiful papers
\cite{Giveon:1998ns,deBoer:1998pp,Kutasov:1999xu}.

For tensionless strings on $AdS$ space, in the absence of an
understanding of the complete theory we cannot conclude that similar
spacetime interpretations will apply. However, in light of the
proposal of \cite{karch1,karch2} which suggests a holographic relation
between bosonic strings on zero radius $AdS_{d+1}$ and free field
theory in $d$ dimensions, we expect that a spacetime CFT could emerge for
these strings on $AdS_3$. For such a description to exist we must be able
to construct a candidate spacetime conformal algebra with a non-zero
central
charge (the latter being a necessary requirement for a holographic
interpretation as a unitary conformal field theory in two dimensions).
Below we show that tensionless strings on $AdS_3$ seem to naturally
provide
such a structure.

We propose that the generators
of one copy of the spacetime Virasoro algebra are the time-independent
quantities
on the worldsheet given by
\EQ{L_n\;:=\;\;-i\int_0^1d\sigma \;p_-\,X^{n+1} \;\;=
\;\;-i\int_0^1d\sigma \;p_-\,
{\left[ x^-+{z\dot z\over 2p_-} \right]}^{n+1},\label{ellen}}
for all integers $n$ (for $n<-1$ we need to be careful in defining $L_n$
as $X(\sigma)$ can have zeros). Strictly speaking these
are the classical expressions. In the quantum theory we need to adopt
an ordering prescription to define these operators (we will address
this in Appendices A and B). We first note certain
important properties of these objects.
By construction, they are time-independent and commute
with the light-cone string Hamiltonian since $X(\sigma)$
(Eq.\eqref{bigx}) is time-independent as a consequence of the
constraints.

Using the canonical commutators, it can be easily checked
that the three elements
\AL{L_1 \; =&\; -i \int_0^1 d\sigma\;\left[p_-\,(x^-)^2 + x^- z\dot z +
{1\over 4 p_-}\;z^2 {\dot z}^2\right] \; ,\\\nonumber\\
    L_0 \; =&\; -i\int_0^1 d\sigma\; \left[p_- \, x^- + {1\over 2}
z{\dot{z}}\right] \; ,\\\nonumber\\
    L_{-1} \; =&\; -i p_- \; ,}
generate an $sl(2,{\mathbb R})$ algebra. Again, these are classical
expressions and are subject to a specific ordering in the
quantum theory. At the classical level, closure of the $sl(2,{\mathbb R})$
algebra
follows from canonical Poisson bracket relations. As we discuss in
Appendix B, it is straightforward to check closure on appropriately
defined physical states (defined via Eq.\eqref{physical}) in the quantum
case
as well. The $SO(2,2)\cong {SL(2,{\mathbb R})} \times {SL(2,{\mathbb R})}$
isometry group of $AdS_3$ acts as a global symmetry on worldsheet fields.
$L_{1}$, $L_{0}$ and $L_{-1}$ are the
conserved charges associated with one of these two $SL(2,{\mathbb R})$
invariances.

That $\{ L_n \}$ should actually be thought of as spacetime charges in
$AdS_3$ emerges when we expand out
Eq.\eqref{ellen} as a binomial series (we may think of this as a small $z$
expansion), to obtain
\EQ{L_n = -i\int _0^1
  d\sigma\;\left[p_- \, (x^-)^{n+1}+{1\over2}(n+1)(x^-)^{n} {z\dot z}
+{1\over 8 p_-} \, n(n+1) (x^-)^{n-1} z^2{\dot z}^2 +\ldots\right].}
Recall that $p_-$, $\pi_+ = \dot {x}^- =-{\dot
  z}^2/2p_-$ and $\pi_z=\dot z$ are the canonical momenta conjugate to the
coordinates
$x^-$, $x^+$ and $z$ respectively. Hence the leading terms in an
expansion near the $AdS_3$ boundary become
\EQ{L_n=-i\int_0^1 d\sigma\; \left[ p_- \,(x^-)^{n+1} +{1\over
      2}(n+1)(x^-)^{n}\;z\;\pi_z-{1\over 4}
    n(n+1)(x^-)^{n-1}\;z^2\;\pi_+ +\ldots\right].\label{asympln}}
In Poincar\'{e} coordinates, these are precisely the
asymptotic isometry generators of $AdS_3$, realised on string
worldsheet fields. Recall that one set of asymptotic Killing
vectors of $AdS_3$ are
\SP{&\xi^+_n \; =\; -{1\over 4}\;n(n+1)(x^-)^{n-1}\,z^2\;+\;{\cal O}(z^4)
\; ,\\\
&\xi^-_n \; =\; (x^-)^{n+1}\;+\;{\cal O}(z^4) \; ,\\\
&\xi^z_n \; =\; {1\over 2}\;(n+1)\;(x^-)^n\,z\;+\;{\cal
O}(z^3) \; ,\label{BH}}
for all integers $n$ (in Poincar\'{e} coordinates). It is easily seen that
these vectors satisfy a Virasoro
algebra while the central term in the algebra was otained after a very
careful analysis in \cite{brown}.   It is interesting that the
infinite set of time-independent quantities $\{ L_n \}$, on the
worldsheet theory of the tensionless string,
naturally have a geometric interpretation (in the $z \rightarrow 0$ limit)
in terms of
the asymptotic isometry generators of $AdS_3$. This relation to the
Brown-Henneaux Killing vectors applies only to the first three
terms in the binomial expansion of Eq.\eqref{ellen} given in
Eq.\eqref{asympln}. It is not clear
to us whether the remaining terms have a natural geometrical
interpretation. In light of the above, it is
natural to think of $\{ L_n \}$ as spacetime charges.

Although we have established a link between the worldsheet $\{ L_n \}$ and
the
Brown-Henneaux generators, it is by no means obvious that the former
constitute a Virasoro algebra. However, this fact follows from the
highly nontrivial commutation relation obeyed by
the worldsheet field $X(\sigma)$. This is relatively
easy to establish at the classical level via Poisson brackets, since
\SP{[L_m,L_n]_{PB}= -\int_0^1 d&\sigma\int_0^1 d\sigma^\prime \;
\Big\{(m+1)(n+1)\;p_-^2\;\;[X(\sigma),X(\sigma^\prime)]_{PB}\;\;X^m(\sigma)\;
X ^n(\sigma^\prime)+\\
&+i(m+1)\;p_-\;X^{m}(\sigma)\;X^{n+1}(\sigma^\prime)
-i(n+1)\;p_-\;X^{n}(\sigma^\prime)\;X^{m+1}(\sigma)\Big\}}
where we have used the fact that $X(\sigma)$ and $p_-$ are canonically
conjugate. Using the relation
Eq.\eqref{xcom} (understood as a Poisson bracket) and imposing the
periodicity
of the closed string fields ({\it i.e.} $X(1)=X(0)$) we perform one
of the integrals and find (ignoring the $c$-dependent term) that indeed
\EQ{[L_m,L_n]_{PB}\;=\;(m-n)\;L_{m+n}.\label{bh1}}
For the quantum operators this calculation gets
complicated very rapidly with increasing values of $m$ and $n$ as one
needs to
perform several reorderings that involve evaluating multiple
commutators. However, it is possible to organise the calculation in a
systematic expansion in powers of $1/p_-$, each
order in the expansion being associated with the number of
commutators taken. In this expansion one can then check for
closure of the $\{ L_n \}$ algebra order by order in $1/p_-$ .
We demonstrate closure on physical states at next to leading order
$(p_-)^0$ in Appendix C, the classical result being the leading term of
order $(p_-)^1$.

Up to this point we have ignored the effect of the $c$-dependent term in
Eq.\eqref{xcom} in the algebra. The inclusion of this
terms modifies the algebra as follows
\EQ{[L_m, L_n] = (m-n)L_{m+n}-{c\over 2}(m^2+m)(n+1){1\over 2\pi i}\int
  _0^1 d\sigma\;{X^{m+n-1}(\sigma)\partial_\sigma
X(\sigma)}.\label{undef}}
For $m+n\neq 0$ the second term clearly vanishes as an integral of a total
derivative, due to periodicity
of the field $X(\sigma)$ on the closed string. However, when $m+n=0$,
the integral reduces to $\int d\ln X$ which, depending on its
definition, is either a phase or zero. Note that when $m+n=0$, the
contribution from this term is proportional to $(m^3-m)$, as expected
for a central extension.
(Note also that the factor of $1/2\pi
i$ that naturally appears in Eq.\eqref{undef}, as a consequence of the
commutation relation Eq.\eqref{xcom}, ensures that this term is a
real integer). The definition of this object is related to
the definition of the generators $\{ L_n \}$ for $n<-1$. However, we can
understand its interpretation by following a slightly
different route.

Let us consider strings on $AdS_3$ in global
coordinates \cite{review}. Although we cannot directly formulate the
worldsheet theory at zero radius in these coordinates, we can do so near
the
boundary. The boundary of $AdS_3$ in global coordinates is a
cylinder and the asymptotic metric (near the boundary) is
\EQ{{\rm lim}_{\rho\rightarrow\infty} \;\; ds^2 \;\simeq\;
  R^2(\,2\,e^{2\rho}d\tilde x^+d\tilde x^- + d\rho^2 \, ),}
where $\rho$ is the radial coordinate and $\tilde x^\pm:=(\phi\pm
\tilde t\;)/\sqrt 2$ with $\phi$ being the angular variable
$0\leq\phi<2\pi$ and $\tilde t$ being the
global time coordinate. With this form of the near boundary metric we can
look at the worldsheet theory in light-cone gauge with
$R^2/\alpha^\prime=0$, after redefining $\exp{(-\rho)}=\tilde z$. The
worldsheet action, constraints and mode expansions are then similar
to those we have already encountered, but with one difference:
since $\phi$ parametrises a circle we can allow for winding
modes of the string and the mode
expansion of $\tilde x^-$ can have a linear term $2\pi w\,\sigma$, with
$w$ an
integer. Furthermore, in these coordinates we need to adopt a slightly
different definition of the
spacetime Virasoro generators, so that they naturally generate
diffeomorphisms of the circle (at the boundary of $AdS_3$). This leads
us to consider generators in an exponential parametrisation
\EQ{H_n :=-\int_0^1 d\sigma\;p_-\;\exp[in \tilde X(\sigma)],\qquad
\tilde X(\sigma):= \tilde x^-+\tilde z\dot{\tilde z}/2p_- .}
$\tilde X(\sigma)$ (in the asymptotc sense discussed
above) obeys the same commutator as
in Eq.\eqref{xcom}. Near the boundary, $H_n$ reduce to the Brown-Henneaux
Killing vectors in global coordinates. From the commutation relations it
then follows that
\EQ{[H_m,H_n]=(m-n)H_{m+n}+{c\over
    2}\,m^3\,\delta_{m+n,0} {1\over 2\pi}
\int_0^1\;d\sigma\;\partial_\sigma\tilde
  X (\sigma).}
The presence of a linear term $2\pi w\,\sigma$ in ${\tilde x^-} ( \sigma
)$
leads to
an unambiguous central extension $(c/2)w\, m^3 \,\delta_{m+n,0}$. We
expect the theory on $AdS_3$ in global coordinates
to be dual to a CFT on the cylinder (or radially quantized CFT), whilst
the theory in Poincar\'{e} coordinates (discussed previously) should be
dual to a CFT
on the plane. For the radially quantized theory, the central term above
suggests how we should interpret the logarithm in $\int d\ln X$ in
Eq.\eqref{undef}. Noting that, asymptotically, the global coordinate
$\tilde x^-$ is related to the Poincar\'{e} coordinate $x^-$ such that
$x^-=\tan{\tilde x^-}$, we must define the logarithm so that it picks out
the appropriate phase and the
central charges in the two descriptions match\footnote{The
  logarithmic integral also yields a phase if we add a small imaginary
  part to $X(\sigma)$, in our definition of $\{ L_n \}$, whose sign
depends
  on how we approach a zero of $X( \sigma )$.}. This leads to the
following
algebra for $\{ L_n \}$:
\EQ{[L_m, L_n] = (m-n)L_{m+n}+{c\over
    2}w(m^3-m)\delta_{m+n,0}}
where $w$ is a \lq winding' number defined via the logarithmic integral,
as explained above. Thus, from the
point of view of the worldsheet theory of the tensionless string on
$AdS_3$, we can see the emergence of a spacetime Virasoro algebra
with a potential central extension \footnote{J. Troost has pointed out
to us that one can obtain an unambiguous definition of the
central term directly in Poincar\'{e} coordinates if one imposes an 
angular identification of the coordinate $x$ in $AdS_3$. With this
identification $AdS_3$ becomes the 
vacuum BTZ black hole and the spacetime central
charge arises from winding states in this geometry (for a discussion
of the tensile
string in this geometry see \cite{Troost:2002wk}and references therein).}

. This mechanism indicates that a
holographic
interpretation as a boundary CFT should be possible for the zero radius
(or
tensionless) theory (e.g. as would follow from a naive extension
of the proposal of \cite{karch1,karch2}). A similar
mechanism for generating a spacetime central charge was found in a
very different situation in the context of the $SL(2,\mathbb R)$ WZW
model in \cite{Giveon:1998ns} where the winding of long strings in
$AdS_3$ induced a central charge in the spacetime Virasoro algebra.
(A different mechanism by which a spacetime central charge is also
induced in the $SL(2,\mathbb R)$ WZW model was shown to operate in
\cite{deBoer:1998pp}.)

\subsection{A second copy of the Virasoro algebra}

Having argued for the existence of an infinite-dimensional spacetime
conformal symmetry (generated by two copies of a Virasoro algebra), we now
construct the second copy of the Virasoro algebra.
Following a similar approach as
before, we first identify two time-independent worldsheet fields
\EQ{Y(\sigma)\;:=\; x^+ (\tau )+ {1 \over 2 \pi_+ (\sigma)} z{\dot
    z} (\sigma , \tau ) ,\qquad
\pi_+(\sigma)\;=\;-{1\over 2p_-}{\dot z}^2 (\sigma),\label{bigy}}
where $x^+ (\tau) = p_-\tau$. It is easily verified that the two variables
$Y$ and $-\pi_+$ are \lq
conjugate', in the sense that they satisfy the simple commutation
relation,
\EQ{[Y(\sigma),
  \;-\pi_+(\sigma^\prime)]\;=\;i\delta(\sigma-\sigma^\prime).\label{ycom}}
We now propose that the second set of Virasoro generators are
\EQ{\tilde L_n \; :=\; i\int_0^1 d\sigma\;\pi_+(\sigma)\; Y(\sigma)^{n+1}
\label{ellentilde}}
which can be shown to commute with each $L_n$,
\EQ{[L_m,\tilde L_n]=0.}
Again, a small $z$ expansion of these conserved quantities yields the
following leading terms
\EQ{\tilde L_n= i\int_0^1d\sigma\; \left[\pi_+ \,{(x^+)}^{n+1}+
{1\over 2}\,(n+1)\,(x^+)^n\,z\,\pi_z -{1\over
  4}\,n(n+1)\,(x^+)^{n-1}\, z^2 \, p_-+\ldots\right]}
which agree with the other set of the asymptotic Killing vectors of
$AdS_3$, namely
\SP{&\tilde \xi^+_n \;=\; (x^+)^{n+1}\;+\;{\cal O}(z^4) \; ,\\\
&\tilde\xi^-_n \;=\;-{1\over 4}\;n(n+1)(x^+)^{n-1}\,z^2\;+\;{\cal
O}(z^4) \; ,\\\
&\tilde\xi^z_n \;=\;{1\over 2}\;(n+1)\;(x^+)^n\,z \;+\;{\cal
  O}(z^3) \; ,}
for all integers $n$. That $\{ \tilde L_n \}$ constitute a Virasoro
algebra follows from
the fact that $-\pi_+$ and $Y$ are canonically conjugate in the sense of
Eq.\eqref{ycom}. This is easy to establish classically, at the level of
Poisson
brackets. For quantum operators two important facts need to be taken
into account. First, there is the usual issue of operator
ordering. Second, the algebra actually acquires a central
extension via the same mechanism encountered in the previous section in
the context of the algebra of $\{ L_n \}$. To see the emergence of this
central
extension it is convenient to adopt a slightly indirect approach. We
first note that the variables $X$ and $Y$ are related by the following
relation,
\EQ{p_-X^\prime(\sigma) \;=\; \pi_+(\sigma)Y^\prime(\sigma).}
Now the Fourier components of $p_- X^\prime$ (and hence those of
$\pi_+Y^\prime$) are the generators of the worldsheet Virasoro
algebra Eq.\eqref{vir2}, so that
\EQ{[\pi_+(\sigma) Y^\prime(\sigma),\pi_+(\sigma^\prime)
  Y^\prime(\sigma^\prime)]\;= \;i\pi_+
    Y^\prime\left({\sigma+\sigma^\prime \over2}\right)
\delta^\prime\left({{\sigma-\sigma^\prime}\over 2}\right)-{c\over
  2}\,{1\over 2\pi i}
\,\delta^{\prime\prime\prime}(\sigma-\sigma^\prime).}
This relation implies that $[Y(\sigma),Y(\sigma^\prime)]\sim c\;
\delta^\prime(\sigma-\sigma^\prime)/{\pi_+(\sigma)}\pi_+(\sigma^\prime)$,
leading to a central term in the spacetime Virasoro
algebra. Following a short calculation, similar to that in the
previous section, we find
\EQ{[\tilde L_m,\tilde L_n]\; =\;(m-n)\;\tilde L_{m+n}\;+ \;{c\over
2}\,\tilde
  w\,(m^3-m)\,\delta_{m+n,0}}
where $\tilde w = \int _0^1 d\sigma \partial_\sigma \ln Y$ is the
\lq winding' associated with the worldsheet field $Y(\sigma)$.

$L_1$, $L_0$ and
$L_{-1}$ and $\tilde L_1$, $\tilde L_0$ and $\tilde L_{-1}$ generate
the $SL(2,{\mathbb R})\times SL(2,{\mathbb R})$ isometry group of $AdS_3$
and are the
conserved charges on the worldsheet corresponding to associated
global symmetries. As we discuss in Appendix B, this algebra can be
shown to close (on
physical states) in the quantum theory as well.

\subsection{Interpretation via worldsheet symmetries}

Although we have argued above that $\{ L_n \}$ and $\{ \tilde L_n \}$
constitute an infinite-dimensional conformal algebra on zero radius
$AdS_3$ (and demonstrated this by explicit evaluation of commutators),
we have not addressed the question of why these conserved
quantities realise Virasoro algebras in the first place. Put
another way, we know that in general on the worldsheet, the isometry
generators of
the target space are realised as conserved charges associated with global
worldsheet symmetries. (Actually, in light-cone gauge the
worldsheet symmetries are combinations of global transformations and
compensating worldsheet reparametrisations that preserve the gauge
choice.)
This is certainly true for the generators of the
$SL(2,{\mathbb R})\times SL(2,{\mathbb R})$ isometry group of $AdS_3$. It
is
natural to ask if the same is true for the infinite sets $\{ L_n \}$ and
$\{ \tilde
L_n \}$. We now argue that this is indeed the case.

Consider the transformations
\AL{&\delta x^-\;=\;F(X) - {1\over 2p_-}\,z\dot{z}\,F^\prime(X) \; ,\\
&\delta z\;=\;{1\over
  2}\,z\,F^\prime(X)}
where $F$ is an arbitrary function. Treating these as "off-shell"
transformations, $\it{i.e.}$ where we neither make use of the equations of
motion nor the constraints, we see that they leave the
transform
the constraint equation \eqref{vc1} covariantly. They also transform the
second constraint \eqref{vc2} covariantly after using the equation of
motion.
The infinite set of
generators of these symmetries, labelled by functions $F_n (X) :=
X^{n+1}$, indeed
constitute a Virasoro
algebra (as can be checked by explicit calculation). The Noether charges
for these
transformations can then be formally derived by varying the light-cone
action Eq.\eqref{lc2} with the constraints implemented via Lagrange
multipliers. Evaluating them on-shell we find that the charges are
precisely $\{ L_n \}$ (Eq.\eqref{ellen}) (a given charge $L_n$ follows
from the above symmetry with function $F_n$). The
other set of Virasoro generators $\{ \tilde L_n \}$ correspond
to symmetries under the transformations
\AL{&\delta x^-\;=\;-{1\over 2p_-^2}\,{\dot z}^2\,G(Y)-{1\over 2p_-}\,
z\dot{z} \,G^\prime(Y)\\
&\delta z\;=\;{1\over
  2}\,z\,G^\prime(Y)+{1\over p_-}\,\dot{z}\,G(Y),}
where $G$ is an arbitrary functon. The associated charges are $\{
\tilde L_n \}$ (Eq.\eqref{ellentilde}), with each $\tilde L_n$
following from the above symmetry with function $G_n (Y) := Y^{n+1}$.

In a near boundary (or small $z$) expansion, the transformations above are
found to be a combination of Brown-Henneaux spacetime diffeomorphisms and
worldsheet reparametrisations that preserve the light-cone
gauge. Such a correspondence is to be expected since the Brown-Henneaux
transformations (combined with
compensating worldsheet reparametrisations) are only asymptotic symmetries
of
the string action, whilst the transformations above are exact symmetries.

\section{Discussions}
In this note we have shown that the worldsheet theory of bosonic
strings in light-cone gauge on $AdS_3$ at zero radius possesses symmetries
generated by two copies of a
Virasoro algebra. We have shown that the conserved charges on the
worldsheet, associated to these symmetries, can naturally be interpreted
as generators of a spacetime Virasoro algebra. This interpretation is
prompted primarily by the observation that, near the boundary of
$AdS_3$, the conserved charges are isomorphic to the
Brown-Henneaux diffeomorphisms that generate asymptotic isometries of
$AdS_3$.
We also find, from the worldsheet viewpoint, that this
spacetime algebra can pick up a central extension indicating a
potential
holographic interpretation as a dual conformal field theory in two
dimensions. This is consistent with the proposal of
\cite{karch1,karch2}, that bosonic string theory on zero radius
$AdS_{d+1}$ should be dual to free scalar field theory in $d$
dimensions. The appearance of the spacetime Virasoro algebra is
intimately related to a {\it worldsheet} Virasoro algebra satisfied by
an infinite set of conserved quantities on the worldsheet of the
tensionless string. The central extension in spacetime which is
generated by the winding of a worldsheet field is also directly related to
the central charge in the worldsheet algebra. These
features are similar to those encountered in a different physical
context: that of strings on $AdS_3$ with an $NS$-$NS$ two form gauge
field, which can be described as an $SL(2,{\mathbb{R}})$ WZW  model.

{\bf Strings on $AdS_3\times S^3\times M^4$}: As mentioned in the
introduction, we can also look at the tensionless limit (or zero $AdS$
radius limit) of IIB string
theory on $AdS_3\times S^3\times M^4$ ($M^4$ being $T^4$ or $K^3$)
with $R$-$R$ three-form flux. For any finite radius, this theory (which
describes the near horizon limit of the D1-D5 brane system) is dual to a
$1+1$-dimensional sigma model whose target space is a
deformation (blowup) of the symmetric product space $(M^4)^k/S_k$ with
$k=Q_1Q_5$ for $Q_1$ D1-branes and $Q_5$ D5-branes. It is likely that the
limit of zero
radius describes the CFT of the symmetric product (at which
point a supergravity description is not valid). It would be extremely
interesting
to understand how this actually works, given that a light-cone
formulation of the superstring on this background (for any radius)
already exists \cite{metsaev1, metsaev2}. The analysis of our paper
pertains only to the bosonic sector of this theory in the tensionless
limit\footnote{
It is worth noting that a naive analysis, based on the Brown-Henneaux
formula for the central charge in the Virasoro algebra: $c=3R/2G_N$
(where $G_N$ is the three-dimensional Newton constant), would suggest
that zero $AdS_3$ radius implies a zero central charge. This naive
expectation clashes with an expectation based on the duality proposal
of \cite{karch1,karch2} or the singular limit of the D1-D5 system.
However, we
are discussing a limit $R^2/\alpha^\prime\rightarrow 0$ wherein it
should be possible to keep $R/G_N$ fixed, as suggested by the $D1-D5$
example.
It is also conceivable that the Brown-Henneaux formula, which is valid
at large radius, could receive corrections in the opposite limit,
{\it i.e.} the small
radius limit.}.

{\bf Virasoro algebras for $AdS_{d}$}: Although we have restricted our
discussion to $AdS_3$, it turns out that the
conserved charges $\{ L_n \}$ and $\{ \tilde L_n \}$ and the Virasoro
symmetries of the tensionless string theory continue to exist for
$AdS$ space of any dimension.
In particular, we can simply extend our definitions of the time-independent
fields $X$ and $Y$ for the theory on
$AdS_{d}$ as
\EQ{X \; :=\; x^-+{1\over 2 p_-} z^i {\dot z^i} \; ,\qquad Y
\; :=\; x^++ {1 \over 2 \pi_+} z^i {\dot z^i}}
with $i=1,...,d-2$. The charges $L_n$
and $\tilde{L}_n$, defined as in Eqs.\eqref{ellen} and
\eqref{ellentilde}, then still satisfy a Virasoro algebra. The
appearance of these symmetries for $AdS_{d}$ suggests a relation
to the generalisation to higher dimensions of the three-dimensional
Brown-Henneaux symmetries, discovered in
\cite{gibbons}\footnote{We
  would like to thank A. Karch for pointing out the possible
  connection to
  \cite{gibbons, Brecher:2000pa}.} \footnote{Geometrically, the reason that these
  asymptotic Brown-Henneaux type isometries of $AdS_{d}$ exist is that
  this geometry admits a foliation with codimension $d-3$ $AdS_3$
  slices \cite{Brecher:2000pa}. Thus these generalised transformations are
  just the ordinary 
  Brown-Henneaux diffeomorphisms acting on a given $AdS_3$ slice (for
  fixed values of the $d-3$ transverse coordinates). We thank
  A. Chamblin for explaining this to us.}. It would be 
interesting to explore this connection further
(We show how this could be made
precise in Appendix 
D.). Another interesting
question is whether these 
symmetries can be extended to the tensile (super-)string on
higher-dimensional $AdS$ spaces (with $R^2/\alpha^\prime \neq 0$).

{\bf Acknowledgements:} We would like thank Hector de Vega, Nick
Dorey, Tim Hollowood, Andreas Karch, Asad Naqvi and
M. Ruiz-Altaba for comments and discussions. We
thank Andreas Karch and Asad Naqvi for carefully reading a draft of
this manuscript and for several useful comments. S.P.K. would like to
acknowledge support from a PPARC Advanced Fellowship.

\startappendix

\Appendix{Operator ordering}

In this appendix we review the issue of defining operators (for example
$P_\sigma$)
in the quantum theory which involve two or more powers of
the modes $z_n$ and $p_n$, and are therefore subject to ordering
ambiguities.
In the context of tensionless strings in
flat space (null-strings), this has been analysed in some detail in
\cite{Gamboa:1989px} where they
focussed on two possible
definitions of quantum operators -- referred to as Weyl $(W)$ and \lq
normal'
ordered $(N)$. Usually, Weyl (or hermitian) ordering is taken to mean the
symmetrization of position and momentum
operators in a given expression so that the resulting operator is
hermitian. In the present
context this ordering turns out to be equivalent
to placing all momentum modes $p_n$ to the
right of all position modes $z_n$. On the other hand, \lq normal' ordering
is a
prescription where all positive frequency modes $z_{n >0}$ and $p_{n \geq
0}$ are ordered to the right of all negative frequency modes. In what
follows we focus on some of the main
consequences of adopting either approach in the context of worldsheet
algebras. We will not atempt to discuss the spectrum of physical
states of the spacetime theory.

The light-cone string Hamiltonian $H_{LC}$ has no ordering ambiguities and
in
terms of Fourier modes is
\EQ{H_{LC}={1\over 2}\sum_{n=-\infty}^\infty p_{-n}p_n.}
For the two types of orderings, we define the vacuum $| 0 \rangle$ of the
light-cone
Hamiltonian via
\EQ{p_n|0\rangle_{W}=0 \; ,}
(for all $n$) for Weyl ordering $(W)$ and
\EQ{p_{n \geq 0} |0\rangle_{N}=0 \; , \qquad z_{n>0} |0\rangle_{N}=0 \; ,}
for normal ordering $(N)$. Both definitions are consistent with
canonical commutation relations
although the Weyl ordered definition is perhaps intuitively obvious.

The worldsheet spatial momentum $P_\sigma$ (the coefficient of the linear
term in
Eq.\eqref{xmode}) is defined as
\EQ{P_{\sigma}^{(W)} ={2\pi i\over p_-} \sum_{n= -\infty}^{\infty}
n\,z_{-n}p_n , \qquad P_{\sigma}^{(N)}={2
  \pi i\over p_-}\sum_{n=1}^{\infty} \left( z_{-n}p_n-p_{-n}z_n \right),}
in the two different schemes. Both definitions annihilate the respective
vacuum state. Finally, for
physical
states of the closed string we must impose
\EQ{P_\sigma^{(N),(W)}|{\rm{phys}}\rangle_{N,W}=0}
in both quantisation schemes which enforces the periodicity of the
worldsheet field $x^-(\sigma,\tau)$.

Given these definitions of the Weyl ordered and normal ordered vacua,
we can
derive the central terms in the worldsheet Virasoro algebras
Eqs.\eqref{vir1} and \eqref{vir2} by following the standard trick of
evaluating the
commutators on the respective vacuum states,
$\langle 0|[l_n,l_{-n}]|0\rangle$. The coefficients $c$ and $c^\prime$
in the central extension can be determined by evaluating
$\langle 0|l_1l_{-1}|0\rangle$ and $\langle 0|l_2l_{-2}|0\rangle$ in
two different ways (directly, and by making use of the Virasoro
algebra itself). We find that
for the Weyl
ordered approach $c=c^\prime=0$ whilst for the normal ordered approach
$c=2$ and taking into account the ordering constant (equal to
$-{1\over 12}$) in $l_0$ we find that
$c^\prime=0$.

\Appendix{$AdS_3$ isometry algebra}
The isometry group of $AdS_3$ is generated by $\{ L_1, L_0, L_{-1} \}$ and
$\{ \tilde L_1,
\tilde L_0, \tilde L_{-1} \}$ whose (classical) expressions can be found
in Eqs. \eqref{ellen} and \eqref{ellentilde}. Here we give explicit
expressions for these generators at the quantum level, in terms of
which their
algebra closes. There are no subtleties involved in the Weyl ordered
approach,
the classical calculation and the quantum calculation are identical
and one does not have to perform any additional reorderings of
expressions after the evaluation of the commutators (note that this
applies only to the $sl(2,{\mathbb{R}})\times sl(2,{\mathbb{R}})$ algebra, not to the entire
Virasoro algebra). For normal
ordering, however the situation is different. The normal ordered
expressions for the generators are
\AL{ L_1=&{i\over p_-}\sum_{n\neq 0}\sum_{m,m^\prime}
\big( {m\over n}+{1\over 2}\big)\big({m^\prime \over n}-
{1\over 2}\big)\;:z_{-n-m}p_mz_{n-m^\prime}p_{m^\prime}:-{i\over
  3}p_-P_\sigma^2\\\nonumber
&-ip_-(x^-_0)^2 -ip_-x_0^-{1\over 2 p_-}\sum_m:z_{-m}p_m:\\\nonumber
&-{i\over 2}x_0^-\sum_m :z_{-m}p_m:
-{i\over 4p_-}\sum_{m,m^\prime}:z_{-m}p_m
z_{-m^\prime}p_{m^\prime}:\\\nonumber
&-{i\over 2}p_-P_\sigma x_0^--{i\over 2}\sum_m:z_{-m}p_m:P_\sigma
-{i\over 2}p_-x_0^-P_\sigma+\\\nonumber
&-i\sum_{n\neq 0}\sum_{m,m^\prime}
{1\over p_-}{m\over n}\big({m^\prime\over n}-{1\over 2}\big)
:z_{-m}p_mz_{n-m^\prime}p_{m^\prime}:\\\nonumber
&-i\sum_{n\neq 0}\sum_{m,m^\prime}
{1\over p_-}{m\over n}\big({m^\prime\over n}-{1\over 2}\big)
:z_{n-m^\prime}p_{m^\prime}z_{-m}p_m:\\
L_0 =&-ip_- x_0^--ip_-P_\sigma/2-{i\over 2}\sum :z_{-m} p_m:\\
L_{-1}=&-ip_-\\
\tilde{L}_1=&-i{p_-\over 2}\sum z_{-m}z_m\\
\tilde{L}_0=&{i\over 2} \sum :z_{-m}p_m:+\;{\rm const}\\
\tilde{L}_{-1}=&{-i\over 2p_-}\sum p_{-m} p_m.}
In the above expressions, normal ordering is denoted $::$ and implies that all the
positive frequency modes of $\{z_n\}$ and $\{p_n\}$ are put to the
right so that they annihilate the vacuum. We also choose to keep
all powers of $P_\sigma$ to the right so that they annihilate all
physical states. It is fairly easy to see that both $sl(2,{\mathbb{R}})$ algebras close individually. However, it requires a long
and tedious calculation to establish that the two $sl(2,{\mathbb{R}})$ algebras
commute with each other -- in particular that $L_1$ commutes with
$\{\tilde L_1,\tilde L_0, \tilde L_{-1}\}$. One useful property that
simplifies the calculation somewhat is that $P_\sigma$
commutes with $\{\tilde L_1,\tilde L_0, \tilde L_{-1}\}$.

\Appendix{Quantum closure of the Virasoro algebra}
In this appendix we demonstrate that the set of conserved charges $\{
L_n \}$ form a
Virasoro algebra on physical states. Instead of discussing the
calculation in its entirety, for clarity, we will just demonstrate the
closure of the commutator algebra at next to leading order (beyond the
leading order classical Poisson bracket calculation). In this context,
"order" means the number of individual commutators that must be
evaluated in reordering the result of a given term in the general $[
L_m , L_n ]$ commutator{\footnote{For example, in the Poisson bracket
    calculation all the terms in each $L_n$ are just treated as fields
    (rather than operators) and can therefore be ordered
    arbitrarily. Hence, after evaluating the first Poisson bracket of
    all such terms with each other, the resulting expressions can be
    reordered at the expense of no further Poisson
    brackets.}}. Keeping track of the order of terms in the
calculation is made easier by the fact that both the commutators $[
X(\sigma ) , X( \sigma^\prime ) ]$ and $[ X( \sigma ) , p_- ]$ reduce
the power of $p_-$ in a given term by one. Since each $L_n$ has just
one power of $p_-$ then this means that we can express $[ L_m , L_n ]
= \sum_{k=1}^{\infty} {({p_-})}^{2-k} {\cal{O}}_k $ for some set of
$p_-$-independent operators $\{ {\cal{O}}_k \}$. Closure of the
algebra then requires that both $p_- {\cal{O}}_1 = (m-n) L_{m+n}$ and
${\cal{O}}_k =0$ for all $k>1$ (up to constraint terms that annihilate
physical states).

In order to verify this statement to order $k=2$, we must first
calculate the $[ X(\sigma ) , L_n ]$ commutator, keeping only leading
order (${( p_- )}^0$) and next to leading order (${( p_- )}^{-1}$)
terms. The result is that
\SP{[ X(\sigma ) , L_n ] \; =\; & \Big( X^{n+1} (\sigma ) +
  \frac{1}{2} ( X^{n+1} (1) - X^{n+1} (0) ) \Big)\\
&+ (n+1) \, X^{\prime} (\sigma ) \int_0^1 d \sigma^\prime \; ( h(
\sigma^\prime - \sigma ) - \sigma^\prime ) \, X^n ( \sigma^\prime )\\
&+\frac{n(n+1)}{2} \frac{i}{p_-} \Big\{ X^\prime ( \sigma ) \int_0^1 d
\sigma^\prime \; ( h( \sigma^\prime - \sigma ) - \sigma^\prime ) \,
X^{n-1} ( \sigma^\prime ) \\
&\;\;\;\;\;\;\;\; +\frac{1}{n} \Big( X^{n} (\sigma ) + \frac{1}{2} (
X^{n} (1) - X^{n} (0) ) - \int_0^1 d \sigma^\prime \; X^n (
\sigma^\prime ) \Big)\\
&\;\;\;\;\;\;\;\; -\int_0^1 d \sigma^\prime \; ( h( \sigma^\prime -
\sigma ) - \sigma^\prime ) \, X^{n-1} ( \sigma^\prime )\\
&\;\;\;\;\;\;\;\;\;\;\;\; \times \Big\{ ( h( \sigma^\prime - \sigma )
- \sigma^\prime ) \, X^{\prime\prime} ( \sigma ) + ( \delta ( \sigma -
\sigma^\prime ) - 1 ) \, X^\prime ( \sigma ) \\
&\;\;\;\;\;\;\;\;\;\;\;\;\;\;\;\;\; + \delta (\sigma - \sigma^\prime )
X^\prime ( \sigma^\prime )  \Big\} \Big\} }%
up to terms involving $X(1) - X(0)=P_\sigma$ ordered to the right (which
annihilates physical states) plus further terms of order ${( p_-
  )}^{-2}$. The function $h$ is defined by $h( \sigma ) := \sigma -
1/2 \;{\rm sgn}( \sigma )$, where ${\rm sgn}(\sigma )$ equals plus or
minus one when
$\sigma$ is respectively positive or negative. The above result then
implies that
\EQ{[ L_m , L_n ] \; =\;  (m-n) \, L_{m+n} }
up to terms involving $X^{r} (1) - X^{r} (0)$ (for various powers $r$)
which we take to annihilate physical states, plus further terms of
order ${( p_- )}^{-1}$. (Note that classically, periodicity on the
closed string implies $X^r(1)-X^r(0)=0$, but quantum mechanically we
can only require that such operators annihilate physical states of the
theory.)
In particular this shows that $p_- {\cal{O}}_1
= (m-n) L_{m+n}$ and ${\cal{O}}_2 =0$ as claimed.

\Appendix{Virasoro algebras for $AdS_{d}$}

Although the main topic of this note is $AdS_3$, it turns out that
straightforward generalisations of the conserved charges $\{ L_n \}$
and $\{ \tilde L_n \}$ also exist in $AdS$ space of any dimension
$d$. We identify two possible generalisations to $AdS_d$. The first is
obtained by defining a time-independent field $X(\sigma)$ as before,
\EQ{X \; :=\; \left( x^- + {1\over 2 p_-} \, z^i {\dot{z}}^i \right)}
where $i = 1,...,d-2$ run over the directions transverse to the
light-cone on $AdS_{d}$ in Poincar\'{e} coordinates. It is a simple
excercise to demonstrate that, with this generalised
$X$, $\{ L_n \}$ (as defined in Eq.\eqref{ellen}) continue to satisfy a
Virasoro algebra. The same is
true for $\{\tilde L_n\}$ (using the obvious generalisation of $Y$). The
existence of these Virasoro algebras for
general $d$ is somewhat mysterious. It has been noted in
\cite{gibbons} that the Brown-Henneaux symmetries can be generalised to
general $AdS_d$ spacetimes. However, the generalised $\{L_n\}$ and $\{\tilde
L_n\}$ above do not match with the generalised Brown-Henneaux
diffeomorphisms in \cite{gibbons}. Hence the geometrical interpretation of
our generalised Virasoro algebra is not entirely clear. Despite this, we are
able to identify another set of conserved quantities on the
string worldsheet which do reduce to the diffeomorphisms in \cite{gibbons},
near the $AdS_d$ boundary.

To construct these conserved quantities, we begin by defining the following
matrix of
time-independent worldsheet fields
\EQ{X^{ij} \; := \; \delta^{ij} \left( x^- + {1\over 2 p_-} \, z^k
    {\dot{z}}^k \right) + {i \over 2 p_-} \sqrt{(d-2) \over 2} \left( z^i
    {\dot{z}}^j - z^j {\dot{z}}^i \right) \; .}
Notice that $X^{ij}$ correctly reduces to $X$ (defined in Eq.\eqref{bigx})
for
$d=3$. The corresponding generalisation of $Y$ is given by the matrix of
conserved quantities $Y^{ij}$, defined just as $X^{ij}$ but with $x^-$ and
$p_-$ replaced by $x^+$ and $\pi_+$ respectively (this definition reduces to
Eq.\eqref{bigy} for $d=3$). We now define the conserved charges
\EQ{{\cal L}_n\; := \; -{i \over d-2} \, \int_0^1 d \sigma \; p_- \, {\rm
   tr}\,\left( X^{n+1} \right) (\sigma) \; ,}
where powers of matrices $X^{ij}$ are taken with respect to the usual
matrix product and the trace is taken by contracting $\delta^{ij}$
with the remaining two free indices. This definition correctly reduces
to the definition of $L_n$ in Eq.\eqref{ellen} for $d=3$. The corresponding
expressions for the conserved charges ${\tilde{\cal L}}_n$ are defined just
as ${\cal L}_n$, but with $X^{ij}$ and $p_-$ replaced with $Y^{ij}$ and
$-\pi_+$ respectively . This also correctly reduces to the definition of
${\tilde L}_n$ in Eq.\eqref{ellentilde} for $d=3$.

Performing a small $z$ expansion of ${\cal L}_n$ gives the following leading
order terms
\EQ{{\cal L}_n=-i\int_0^1 d\sigma\; \left[ p_- \,(x^-)^{n+1} +{1\over
      2}(n+1)(x^-)^{n}\,z^i \,\pi_z^i -{1\over 4}
    n(n+1)(x^-)^{n-1}\,z^i z^i \,\pi_+ +\ldots\right] \; ,}
where we have used the fact that $p_-$, $\pi_+ = \dot {x}^- =-{\dot z}^i
{\dot z}^i /2p_-$ and $\pi_z^i=\dot z^i$ are the canonical momenta conjugate
to the coordinates $x^-$, $x^+$ and $z^i$ respectively. Similarly, an
expansion of ${\tilde{\cal L}}_n$ near the $AdS_d$ boundary yields
\EQ{{\tilde{\cal L}}_n= i\int_0^1d\sigma\; \left[\pi_+ \,{(x^+)}^{n+1}+
{1\over 2}\,(n+1)\,(x^+)^n\,z^i \,\pi_z^i -{1\over
  4}\,n(n+1)\,(x^+)^{n-1}\, z^i z^i \, p_-+\ldots\right] \; .}
In Poincar\'{e} coordinates, these precisely correspond to the asymptotic
isometry generators of $AdS_{d}$ found in \cite{gibbons}.

\end{document}